\documentclass[aps, prb, twocolumn, a4paper, superscriptaddress]{revtex4-2}

\usepackage{graphicx} % Include figure files
\usepackage{dcolumn} % Align table columns on decimal point
\usepackage{amsmath, bm} % maths packages
\usepackage{hyperref} % add hypertext capabilities
\usepackage{color} % colour in document
\usepackage{gensymb} % degree symbol

\makeatletter
\g@addto@macro\bfseries{\boldmath}
\makeatother

\def\crms{Cr$_{1/3}M$S$_2$}
\def\meq{$M$ = Nb or Ta}
\def\crnbs{Cr$_{1/3}$NbS$_2$}
\def\crtas{Cr$_{1/3}$TaS$_2$}
\def\musr{$\mu^+$SR}

\begin{document}

\title{Energy-gap driven low-temperature magnetic and transport properties in \texorpdfstring{\crms}{Cr1/3MS2}\ (\texorpdfstring{\meq}{M = Nb or Ta})}

\author{T.~J.~Hicken}
\altaffiliation{Current address: Department of Physics, Royal Holloway, University of London, Egham, TW20 0EX, United Kingdom}
\affiliation{Centre for Materials Physics, Department of Physics, Durham University, Durham, DH1 3LE, United Kingdom}
\author{Z.~Hawkhead}
\affiliation{Centre for Materials Physics, Department of Physics, Durham University, Durham, DH1 3LE, United Kingdom}
\author{M.~N.~Wilson}
\affiliation{Centre for Materials Physics, Department of Physics, Durham University, Durham, DH1 3LE, United Kingdom}
\author{B.~M.~Huddart}
\affiliation{Centre for Materials Physics, Department of Physics, Durham University, Durham, DH1 3LE, United Kingdom}
\author{A.~E.~Hall}
\affiliation{Department of Physics, University of Warwick, Coventry, CV4 7AL, United Kingdom}
\author{G.~Balakrishnan}
\affiliation{Department of Physics, University of Warwick, Coventry, CV4 7AL, United Kingdom}
\author{C.~Wang}
\affiliation{Laboratory for Muon-Spin Spectroscopy, Paul Scherrer Institut,	Forschungsstrasse 111, 5232 Villigen PSI, Switzerland}
\author{F.~L.~Pratt}
\affiliation{ISIS Pulsed Neutron and Muon Facility, STFC Rutherford Appleton Laboratory, Harwell Oxford, Didcot, OX11 OQX, United Kingdom}
\author{S.~J.~Clark}
\affiliation{Centre for Materials Physics, Department of Physics, Durham University, Durham, DH1 3LE, United Kingdom}
\author{T.~Lancaster}
\affiliation{Centre for Materials Physics, Department of Physics, Durham University, Durham, DH1 3LE, United Kingdom}

\date{\today}

\begin{abstract}
	The helimagnets \crms\ (\meq) have attracted renewed attention due to the discovery of a chiral soliton lattice (CSL) stabilized in an applied magnetic field, but reports of unusual low-temperature transport and magnetic properties in this system lack a unifying explanation.
	Here we present electronic structure calculations that demonstrate the materials are half-metals.
	There is also a gap-like feature (width in range 40--100~meV) in the density of states of one spin channel.
	This electronic structure explains the low-temperature electronic and magnetic properties of \crms\ (\meq), with the gap-like feature particularly important for explaining the magnetic behavior.
	Our magnetometry measurements confirm the existence of this gap.
	Dynamic spin fluctuations driven by excitations across this gap are seen over a wide range of frequencies (0.1~Hz to MHz) with AC susceptibility and muon-spin relaxation (\musr) measurements.
	We show further how effects due to the CSL in \crnbs, as detected with \musr, dominate over the gap-driven magnetism when the CSL is stabilized as the majority phase.
\end{abstract}

\maketitle

Transition metal dichalcogenides ($MX_2$, $M=$ transition metal and $X=$~chalcogen) have long been the subject of much research effort~\cite{dickinson1923crystal}, the reduced dimensionality~\cite{manzeli20172d} of these semiconductors leads to a suite of exciting electrical, optical and mechanical properties~\cite{wilson1969transition,wang2012electronics,choi2017recent}.
More recently, \crms, \meq\ (based on the same structure as superconducting NbS$_2$, but with Cr atoms intercalated between $M$ layers) have attracted attention as rare examples of 2D materials with helical magnetic ordering, thought to occur due to the competition between exchange, the Dzyaloshinksii-Moriya interaction (DMI) and spin-orbit coupling (SOC)~\cite{moriya1982evidence,kousaka2016long}.
Application of a magnetic field perpendicular to the $c$ axis causes the helical ground state to transform to the chiral soliton lattice (CSL)~\cite{togawa2012chiral,zhang2021chiral}, consisting of domain-wall-like 360$\degree$ magnetic kinks separated by regions of ferromagnetic ordering, with a field along a different direction tilting the CSL~\cite{yonemura2017magnetic,mayoh2021submitted}.
These magnetic solitons, reduced-dimensional analogues of skyrmions~\cite{lancaster2019skyrmions}, are topological spin structures with many potential applications~\cite{togawa2012chiral}.
The existence of the CSL in \crnbs\ is well established~\cite{togawa2012chiral}, but has only recently been reported in \crtas~\cite{zhang2021chiral}.

There have been multiple reports of unusual low-temperature transport and magnetic properties in \crnbs, including a sizable change in the thermal conductivity and Seebeck coefficient around $T=40$~K~\cite{ghimire2013magnetic}, and a rapid increase in the Hall coefficient below 50~K~\cite{bornstein2015out}, all suggesting modifications to the electronic behavior.
Additionally, muon-spin rotation (\musr) identifies an increase in oscillation frequency below 50~K~\cite{braam2015magnetic}, suggesting an alteration to the local magnetic-field distribution.
The above observations have led to contrasting explanations, including a change in the dominant transport mechanism due to spin-orbit-coupling-induced alterations of the electronic structure~\cite{ghimire2013magnetic}, or an increased helical length at low $T$ due to a decrease in the magnitude of the DMI~\cite{braam2015magnetic}.
Such explanations do not account for all of the observations, and while an increase in the helical length below 90~K is observed in lamellae~\cite{togawa2019anomalous}, this is not found in bulk~\cite{borisenko2017structural}.
In this Letter we explain the low-temperature behavior of \crms\ (\meq) using electronic structure calculations supported by AC susceptibility and \musr, and demonstrate how the presence of an energy-gap-like feature in the density of states can account for the observed phenomena over a wide range of energy scales.

\begin{figure}
	\centering 
	\includegraphics[width=\linewidth]{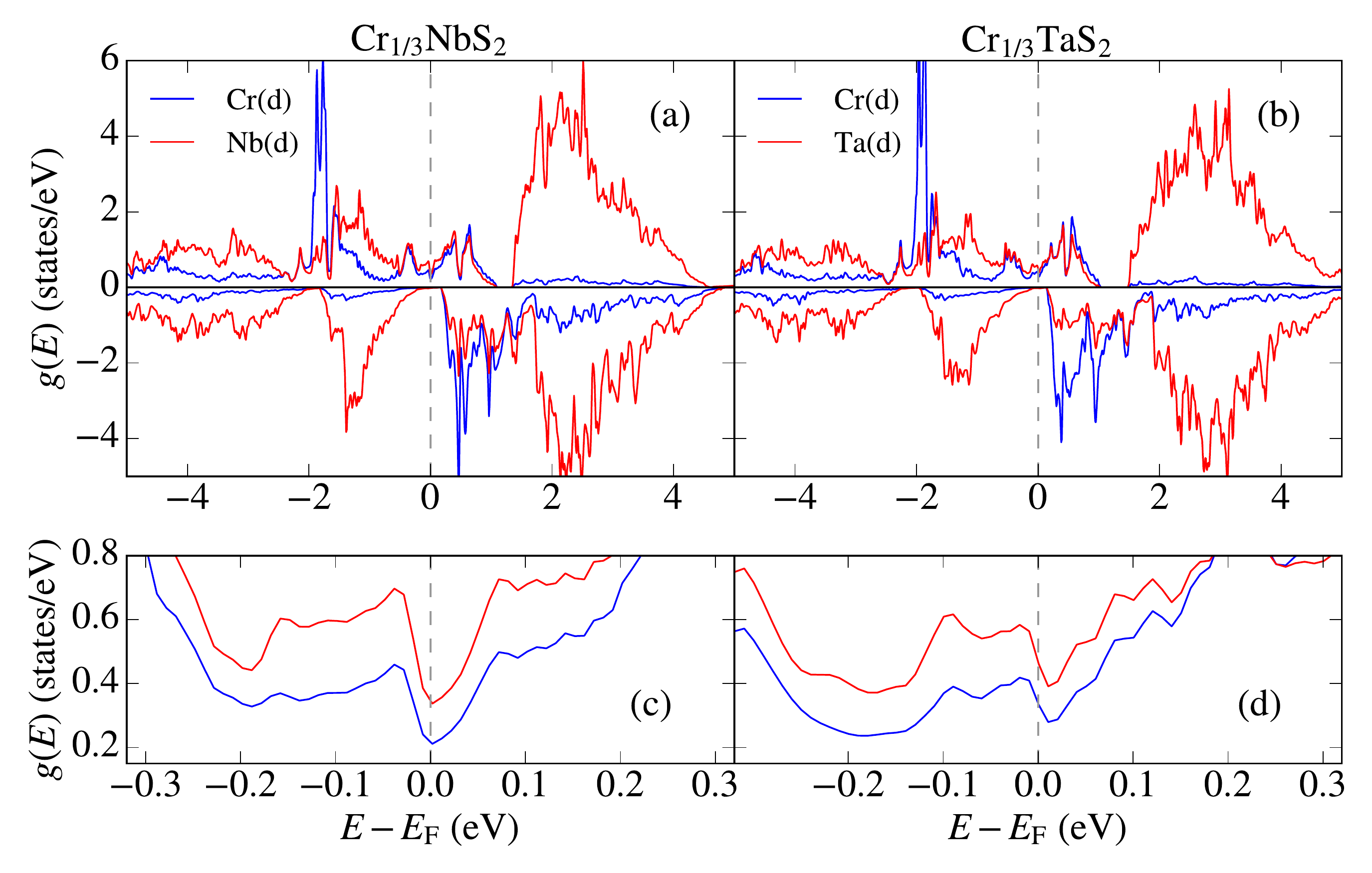}
	\caption{(a,b) DOS of $d$-orbital contributions from transition metals in \crnbs\ and \crtas. (c,d) Regions around the Fermi level showing the pseudogap feature in the spin-up channel. (Spin-up densities are positive and spin-down densities are negative.) }
	\label{fig:bs_and_dos}
\end{figure}

The electronic structure  of \crms\ (\meq) was determined using the \textsc{castep} implementation of spin-polarized density functional theory (DFT)~\cite{clark2005first,perdew1996generalized,monkhurst1976special,sm}.
Spectral calculations were performed to generate band structures and densities of states (DOS) for \crms\ (Fig.~\ref{fig:bs_and_dos}).
In both materials there is splitting of the spin-up and spin-down bands with ferromagnetic moments on the Cr ions.
The largest contribution to the DOS at the Fermi level comes from the spin-up channel, with little contribution from spin-down.
Strongly peaked bands at the $\Gamma$-point in the spin-down channel suggest delocalized electrons; more localized character in the spin-up channel is indicated by flatter bands~\cite{sm}.
The lack of spin-down density around the Fermi level, in addition to the integer spin per unit cell~\cite{coey2004magnetic}, suggests both materials are half-metallic~\cite{coey2002half}.
(Previously \crnbs\ has been reported as a low carrier concentration metal or heavily doped semiconductor~\cite{ghimire2013magnetic}.)
Additionally, in both materials, there is a reduction to almost zero in the spin-up channel of the DOS at the Fermi level with a full-width half-maxima of approximately $\Delta E=80$--$90$~meV.
This reduction has previously been observed in \crnbs\ and termed a pseudogap~\cite{ghimire2013magnetic}, but the influence of this feature was not investigated.
Our calculations performed with the inclusion of SOC have no resolvable impact on the pseudogap~\cite{sm}.

\begin{figure}
	\centering
	\includegraphics[width=\linewidth]{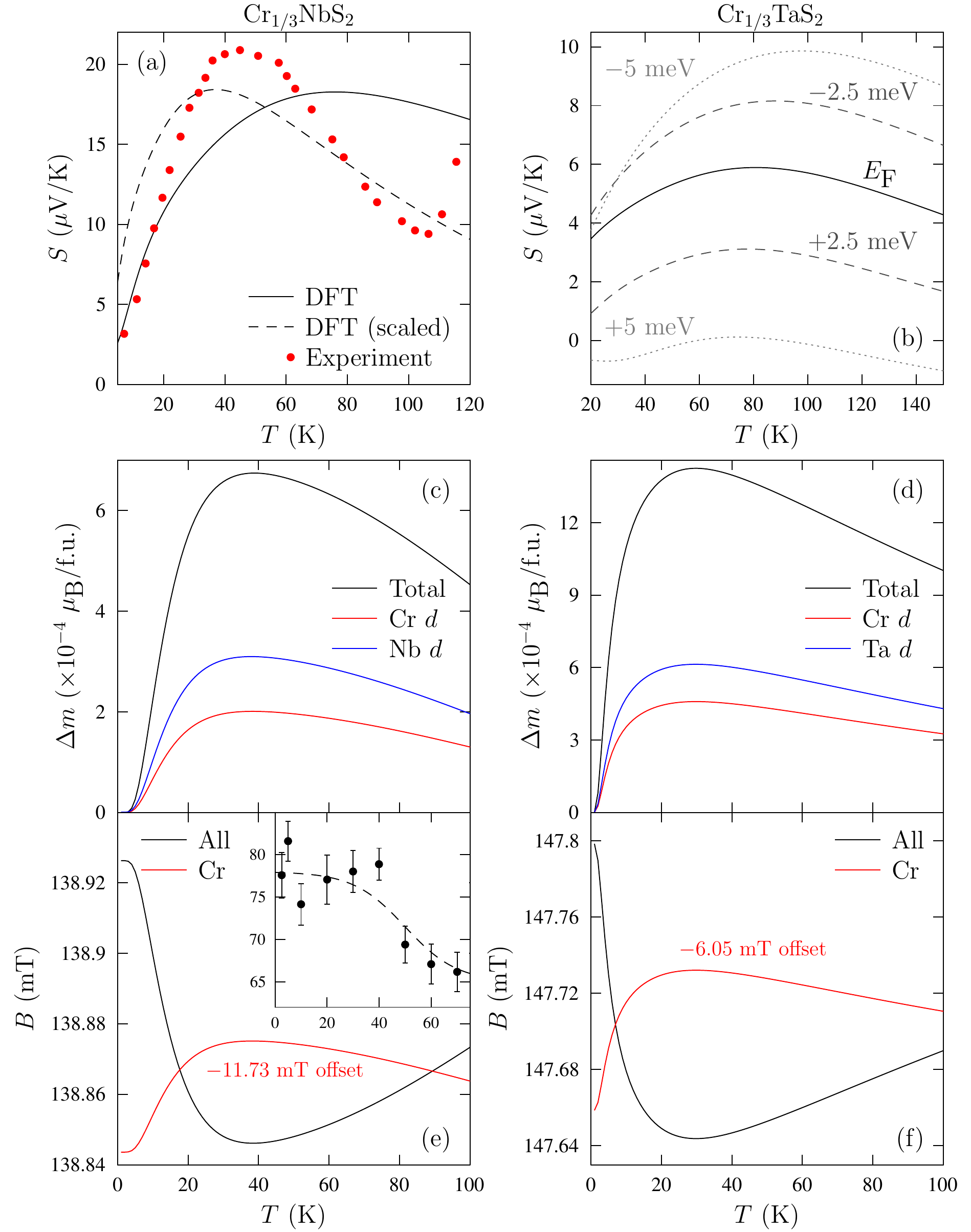}
	\caption{Seebeck coefficient $S=\mathrm{Tr}\left(S_{\mu\nu}\right)/3$ for (a) \crnbs\ and (b) \crtas. In (a) $S$ is compared to experiment~\cite{ghimire2013magnetic}, and is also shown with temperatures scaled according to the gap size (see text). In (b) the range of the possible values of the Seebeck coefficient due to uncertainty in the Fermi energy is shown. (c--d) Temperature evolution of the change in $m$  for \crnbs\ (c) and \crtas\ (d). (Total change, and change in Cr and $M$ $d$-orbitals is shown.) (e--f) Effect on the field at the muon-site in (e) \crnbs\ and (f) \crtas. Field for Cr moment is shown with a vertical offset. Inset: measured field at the muon-site~\cite{braam2015magnetic} [same axes as (e), dashed line is a guide to the eye].}
	\label{fig:dftpred}
\end{figure}

To understand the low-temperature transport properties in \crms\ we calculate the Seebeck coefficient $S_{\mu\nu}$ from DFT~\cite{sm} using the \textsc{boltztrap} code~\cite{madsen2006boltztrap}. Figure~\ref{fig:dftpred}(a--b) shows $S=\mathrm{Tr}\left(S_{\mu\nu}\right)/3$.
This prediction of $S$ for \crnbs\ qualitatively matches the measurements in the low-temperature regime~\cite{ghimire2013magnetic}, although around $T_\text{c}$ the computed behavior deviates from experiment due to the onset of the magnetic transition, which is not captured in our model.
The good agreement between DFT and experiment verifies the calculated electronic structure.
The pseudogap found in DFT is slightly larger than the measured gap (see below), suggesting there are small differences in the energy scale around the Fermi energy.
This results in the features in $S$ to occur at higher $T$ than found experimentally. 
To better compare to the experimental values, we can adjust the energy scale in the calculation to that of the measured gap [$T\rightarrow T\Delta E_\text{exp}/\Delta E_\text{DFT}$, Fig.~\ref{fig:dftpred}(a)], which then provides a better agreement with the experimental peak position.
We find similar behavior of $S$ for \crtas\ [Fig~\ref{fig:dftpred}(b)], as expected from the similar electronic structures.
Our DFT calculations can also reproduce the observed rapid increase in Hall coefficient on decreasing $T$ in \crnbs\ below 50~K~\cite{bornstein2015out,sm}.

The magnetic moment $m$ can be estimated from the difference in the DOS in the spin-up and spin-down channels.
We find, for both materials, $d$-orbitals contribute most significantly to $m$, with around 95\% of the total moment coming from the Cr $d$-orbitals.
We find good agreement between the saturation moment of our calculations and that found experimentally.
The $T$ dependence, $m(T)$, caused by thermally-activated changes in band occupation (but excluding critical behavior due to the phase transition at $T_\text{c}$~\cite{ghimire2013magnetic,mito2019observation}) can be estimated by integrating the DOS weighted by the Fermi-Dirac distribution $f_{\mathrm{FD}}(E,T)$ at a given temperature, $m(T)=\mu_\text{B}\int_{-\infty}^{\infty}\,\left[g^{\uparrow}(E)-g^{\downarrow}(E)\right]f_{\mathrm{FD}}(E,T)~\mathrm{d}E$, where $g^{\uparrow(\downarrow)}(E)$ are the spin-up (spin-down) components of the partial DOS.
The change in $m(T)$ (compared to $T=0$) is shown in Fig.~\ref{fig:dftpred}(c--d).
As $T$ is increased, electrons start to populate excited states, leading to a steady increase in $m(T)$, with the change being dominated by the \meq\ $d$-orbitals [Fig.~\ref{fig:dftpred}(c--d)], despite their small overall contribution to the total moment.

Although these changes in $m(T)$ are likely too small to be detected by a bulk measurement, \musr, a local probe sensitive to the magnitude of the internal field at the muon site, can detect such small changes.
Despite the lack of a magnetic phase transition at low $T$, on the basis of continuous source zero-field (ZF) and transverse-field (TF) measurements, Ref.~\cite{braam2015magnetic} reports an increase in the local field at the muon site [inset of Fig.~\ref{fig:dftpred}(e)].
We calculated the muon stopping site in these materials using DFT~\cite{sm,franke2018magnetic,momma2011vesta} (finding different sites to those determined using nuclear dipole field calculations~\cite{braam2015magnetic}).
The measured field can be compared to predictions using our candidate stopping site to calculate the expected $T$-dependent dipole field at the muon site~\cite{bonfa2018introduction} due to our predicted $m(T)$ [Fig.~\ref{fig:dftpred}(c--d)]~\cite{sm}.
The field at the most probable muon site is highly sensitive to both the amount of Cr intercalated in the material, and to the precise location of the spin density with respect to the muon, neither of which are easy to accurately estimate.
Therefore, our calculation of the absolute magnitude of the field at the muon site, as well as the size of the increase with temperature, does not quantitatively match experiment.
Some of the experimentally observed change may be due to the hyperfine interaction, which typically follows the dipolar field, or due to delocalization of spin; it is not possible to accurately compute these contributions~\cite{sm}.
Despite these limitations, we predict the same trend as found in Ref.~\cite{braam2015magnetic} for \crnbs\ [Fig.~\ref{fig:dftpred}(e)], with a sharp increase in the field at the muon-site at low $T$.
It is notable that the moment of all elements must be considered to obtain this effect; considering Cr alone does not reproduce the trend.

To confirm the influence of the pseudogap on the magnetism, we performed magnetization and AC susceptibility measurements on polycrystalline samples of \crms\ (\meq)~\cite{sm}.
The critical temperature $T_\text{c}$ of each material matches previous reports~\cite{kousaka2016long}, although the field required to induce the forced-ferromagnetic state is lower for both materials~\cite{han2017tricritical}, implying a smaller DMI (likely due to different levels of Cr intercalation affecting the bonding and charge transfer between the $M$ ions, and hence the magnetic properties).
Despite these differences, single-crystal samples of \crnbs\ prepared in the same way as our samples show a CSL in Lorentz transmission electron microscopy measurements~\cite{hall2021unpublished}.

\begin{figure}
	\centering
	\includegraphics[width=\linewidth]{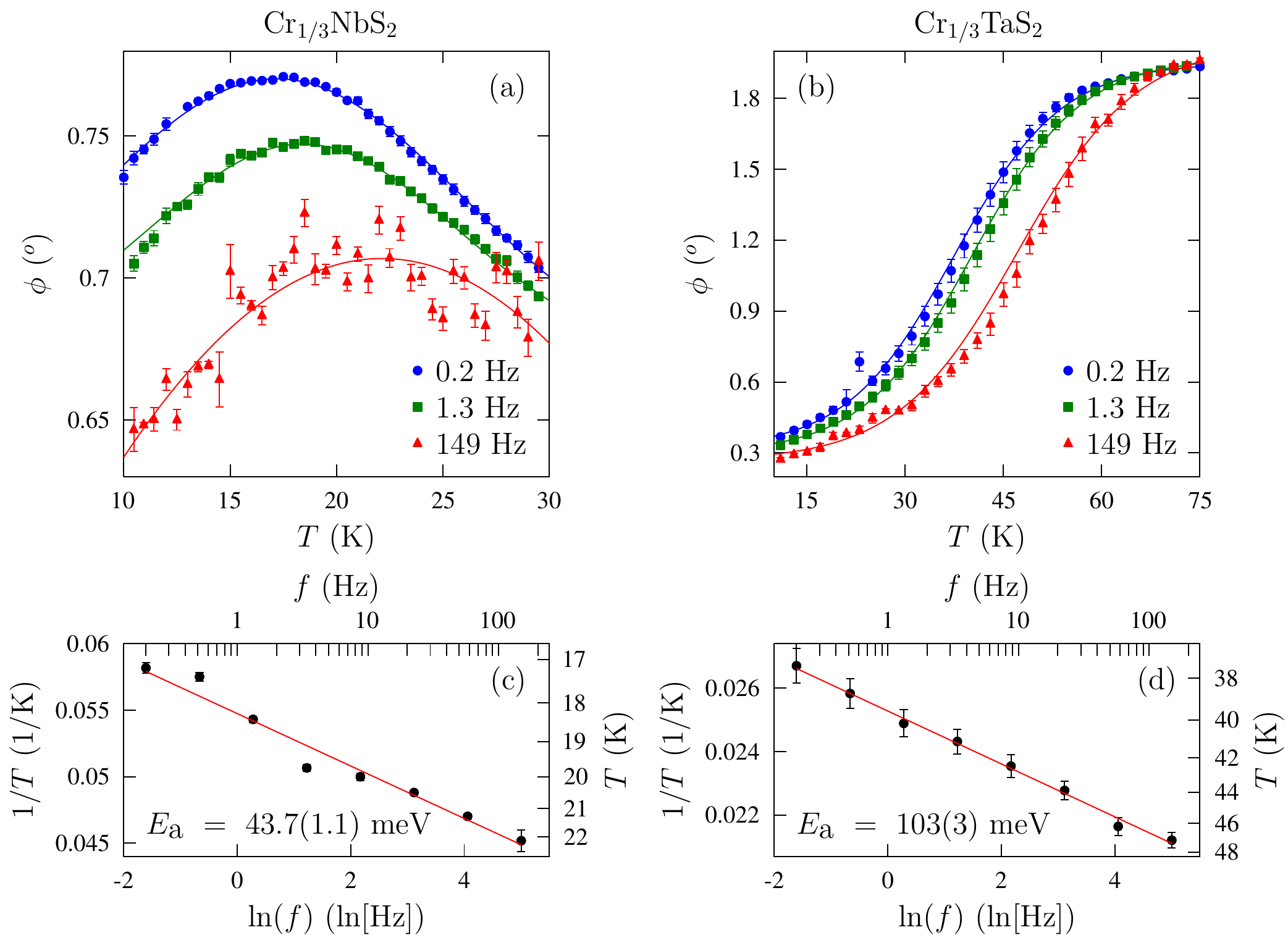}
	\caption{(a--b) Example measurements of the phase lag $\phi$ of AC susceptibility measurements on \crms\ (\meq). Solid lines are fits to the data as described in the text. The central position of these fits is shown in (c--d).}
	\label{fig:phasepeaks}
\end{figure}

Both \crnbs\ and \crtas\ exhibit three regimes of behavior: (i) $T\lesssim T^*$, (ii) $T^*\lesssim T<T_\text{c}$, and (iii) $T>T_\text{c}$, where the low-temperature crossover $T^*\simeq20$~K for \crnbs, and $T^*\simeq45$~K for \crtas.
In region (iii) the materials are paramagnetic, transitioning to long-range order when the samples are cooled into region (ii).
On cooling to region (i), changes are seen most clearly in the phase lag $\phi$ between the applied time-varying magnetic field and the AC susceptibility response.
This suggests dissipative processes~\cite{topping2018ac} determined by fluctuations across the pseudogap.
\crnbs\ shows a peak in $\phi$ around $T^*$, whereas \crtas\ exhibits an increase in this quantity.
For \crnbs\ the behavior is well described by a Gaussian peak with a vertical offset, whereas a sigmoid function with a vertical offset is most appropriate for \crtas.
Both functions allow the extraction of a characteristic temperature $T^{*}$ (the peak location of the Gaussian, or the inflection point of the sigmoid).
Fig.~\ref{fig:phasepeaks}(b,d) shows that $T^{*}$ follows an Arrhenius law, with the gradient giving an activation energy $E_\text{a}$ consistent with the full-width half-maxima of the pseudogap $\Delta E=80$--$90$~meV identified in the spin-up DOS.
[In \crnbs\ the gap is slightly smaller than identified via DFT, and this is the value used to rescale $S$ in Fig.~\ref{fig:dftpred}(a)].
The observed dynamics are therefore consistent with moment fluctuations caused by temperature driven transitions across the gap.

To further probe dynamic fluctuations around the pseudogap, we performed pulsed source longitudinal field (LF) \musr\ measurements on \crms\ (\meq), which are sensitive predominantly to MHz to GHz dynamics (i.e.\ much higher frequencies than the $<150$~Hz dynamics probed with AC susceptibility).
In a LF \musr\ experiment~\cite{blundell1999spin,blundell2021muon,sm} spin-polarized muons are implanted in a sample with the initial muon-spin direction parallel to the externally-applied magnetic field.
Time-varying internal magnetic fields at the stopping site flip the muon spin, reducing the average muon-spin polarization.
The measured quantity of interest is the positron-decay asymmetry, which is directly proportional to the muon-spin polarization.
Both materials show changes in the initial asymmetry $A_0$ near $T_\text{c}$ for all applied fields, indicating the magnetic transition.

\begin{figure}
	\centering
	\includegraphics[width=\linewidth]{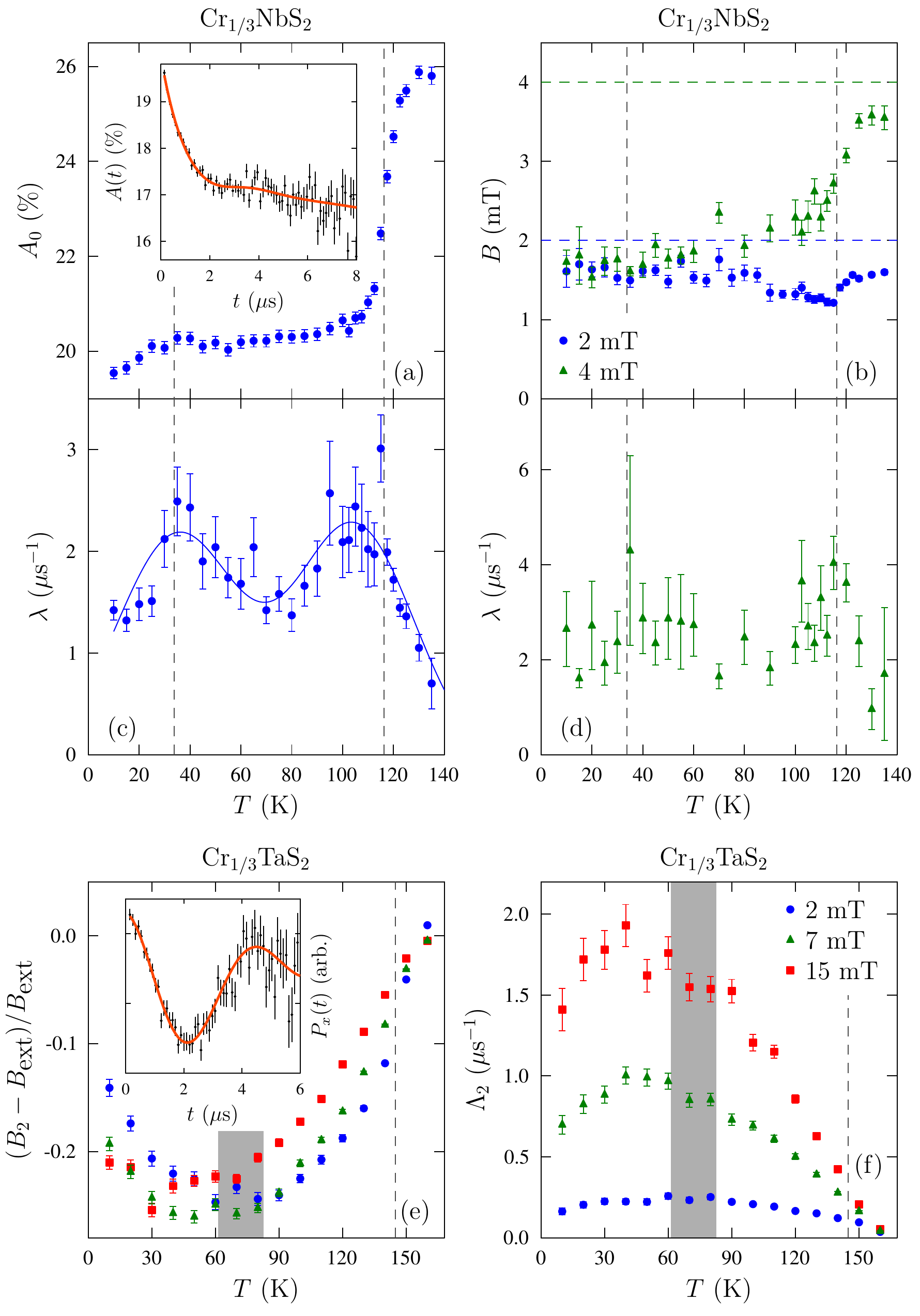}
	\caption{(a--d) LF \musr\ results for \crnbs. (a) Initial asymmetry. Inset shows an example of the spectra and fit for 2~mT and 15~K. Fitted parameters using the model presented in the text are shown in (b--d). The solid blue line in (c) is a guide to the eye. (e--f) Parameters from fitting TF \musr\ measurements of \crtas\ with the model presented in the text. The inset in (e) shows an example of the spectra for 2~mT at 20~K. High $T$ dashed lines indicate $T_\text{c}$, with low $T$ dashed lines indicating $T^*$ (extrapolated from AC susceptibility). For TF \musr\ measurements a gray region indicates $T^*$, accounting for the field dependence.}
	\label{fig:musrmain}
\end{figure}

We first look at \crnbs.
A small applied field will decouple the nuclear contribution to the spectra leaving us sensitive to electronic fluctuations.
An example spectrum for $T<T_{\mathrm{c}}$ is shown in Fig.~\ref{fig:musrmain}, where $\mu_0H_\text{ext}$=2~mT.
A reduction in $A_0$ is observed at low $T$ [Fig.~\ref{fig:musrmain}(a)].
Extrapolating the Arrhenius behavior seen with AC susceptibility to frequencies appropriate for \musr\ ($\simeq\gamma_\mu B_\text{ext}$, see below), gap driven effects should be observed around $T^*=35$~K (marked as a dashed line in Fig.~\ref{fig:musrmain}), and it is below this temperature where changes in $A_0$ are observed.
A reduction in $A_0$ indicates a larger proportion of the muons stopping in sites where there is a component of the magnetic field perpendicular to the initial muon-spin, causing precession too rapid to be observed in the experiment.
This suggests an increase in the internal field at low $T$, as seen with ZF \musr~\cite{braam2015magnetic} in Fig.~\ref{fig:dftpred}(e).
As the external field is increased, the internal field has less influence on the field at the muon site, which is reflected in the reduced resolution of this effect in $A_0$~\cite{sm}.

The LF \musr\ spectra measured on \crnbs\ at all applied fields were fitted~\cite{pratt2000wimda} with $A\left(t\right)=a_1\exp\left(-\sigma^2t^2\right)\cos\left(\gamma_\mu Bt\right)+a_2\exp\left(-\lambda t\right)+a_\text{b}\exp\left(-\lambda_\text{b}t\right)$, where the first term accounts for muons in sites where the average internal field is not parallel to the initial muon-spin, leading to precession in the effective local field $B$, the second term accounts for muons in a purely longitudinal average magnetic field, or those where the static field largely cancels, leading to relaxation from dynamic fluctuations, and the third term accounts for muons that stop outside the sample.
This model should only be applicable below $T_\text{c}$, however we find it describes the data well over the entire temperature range since it also approximates the expected LF Kubo-Toyabe behavior above $T_\text{c}$ that occurs due to relaxation from disordered electronic moments.
The two regimes are separated by a peak in $\lambda$ around $T_\text{c}$.
We find at each field, $\lambda_\text{b}$ and $\sigma$ are $T$-independent, therefore we fix them to $\lambda_\text{b}=0.005$~$\mu$s$^{-1}$ and $\sigma=0.28$, $0.55$, $0.52$~$\mu$s$^{-1}$ for 2, 4, 10~mT respectively.
The $T$-independence of $\sigma\propto\sqrt{\langle\left(B-\langle B\rangle\right)^2\rangle}$ suggests the distribution of magnetic fields at the muon site does not significantly change.
Fitted parameters extracted from the 2~mT measurements are shown in Fig.~\ref{fig:musrmain}(b--d).

The main result from our \musr\ measurement of \crnbs\ is that the observed MHz dynamics are consistent with the pseudogap-driven magnetism.
As shown in Fig.~\ref{fig:musrmain}(c), there is a clear peak in $\lambda$ at $T^*$ despite no reported change in magnetic structure~\cite{borisenko2017structural}.
This peak likely occurs due to the same dynamic fluctuations of the moment size as are responsible for the behavior observed in AC susceptibility measurements. 
As $T$ is increased the fluctuation rate $\nu$ increases in frequency, passing through the muon time-window.
In general, the relaxation rate will be enhanced at $T^{*}$ when $\gamma_\mu B\approx \nu\left(T^{*}\right)$~\cite{sm}, where $\gamma_\mu=2\pi\times 135.5$~MHzT$^{-1}$ is the gyromagnetic ratio of the muon.
The observed peak coincides with the prediction from the Arrhenius behavior of the AC susceptibility when $\nu\left(T^*\right)\approx\gamma_\mu B_\text{ext}$.
This is consistent with the muon sites most sensitive to dynamics being those where the effective field is close to the applied field (i.e. the average static field is small).
This interpretation is also supported by the fitted value of the field $B$ [Fig.~\ref{fig:musrmain}(b)]~\cite{sm}.

We have performed additional LF \musr\ measurements in applied field of 4~mT (where we expect a significant volume fraction of the sample in the CSL state), and 10~mT (where the solitons in the CSL have large separations).
Measurements at these fields were found to be well described by the same model as discussed above.
At 10~mT, the behavior of all fitted parameters was found to be very similar to those at 2~mT (see Supplemental Material~\cite{sm}), however more significant changes are observed at 4~mT.
Most notably, the dynamic changes at $T^*$ observed through $\lambda$ are not seen at 4~mT [Fig.~\ref{fig:musrmain}(d)], but are recovered at 10~mT.
It is possible that since the topologically-protected CSL has a large energy barrier preventing changes to the static structure, the more robust magnetism in this phase might prevent the effects of energy-gap driven magnetism being measurable.
Other differences are seen in the effective local field $B$ [Fig.~\ref{fig:musrmain}(b)], which at 2~mT remains fairly constant at slightly less than the applied external field (with a deviation at $T_\text{c}$).
However, at 4~mT there is a sizable reduction compared to the applied field, with $B$ decreasing continuously with decreasing temperature, perhaps suggesting a continuous evolution in the size of the static field at the muon site in the CSL.
Below $T_\text{c}$ at 10~mT, where the CSL is expected to occur with reduced volume fraction, $B$ is approximately constant at a value significantly reduced compared to the applied field~\cite{sm}.

We have also performed continuous source ZF \musr\ measurements on \crms\ (\meq).
In both cases a fast relaxation of $A\left(t\right)$ is observed at early times, consistent with the oscillations detected in single crystal samples of \crnbs~\cite{braam2015magnetic}, however full oscillations are not resolved in our polycrystalline samples, likely due to the different, complex field distribution arising from a combination of the domain and magnetic structures.
Continuous source TF \musr\ measurements, performed on the same \crnbs\ sample, show very similar early-time behavior to the LF measurements (as expected for an ordered magnet with a large internal field compared to the applied field).
Pulsed source LF \musr\ measurements of \crtas\ indicate the field at the muon site is considerably larger than in \crnbs, hence TF \musr\ measurements are most useful.
To analyze our continuous source TF \musr\ measurements, the number of counts in each detector is simultaneously refined to a single $P_x\left(t\right)$, with an appropriate phase offset applied for the detector geometry.
In these measurements we find $P_x\left(t\right)=\sum_{i=1}^2a_i\cos\left(\gamma_\mu B_it+\phi_i\right)\exp\left(-\Lambda_it\right)$ well describes the data.
The first component captures muons that stop in sites predominantly sensitive to the large internal field and get rapidly relaxed out the spectra.
In this discussion we focus on the second component [Fig.~\ref{fig:musrmain}(d--e)], capturing muons that are predominantly sensitive to the applied field and a small internal field (it is these sites that are most similar to those to which we are sensitive in the LF \musr\ measurements).
We find $\phi_2\simeq0$, as expected for this class of muon sites.
We see an enhancement of the field at the muon site $B_2$ below $T^*$, consistent with similar measurements of \crnbs\ in Ref.~\cite{braam2015magnetic}.
At high $T$, $\Lambda_2$ appears to scale with the average moment as one would expect, however the behavior of this parameter is more complicated below $T^*$.
In this configuration, $\Lambda_2$ is sensitive to both the width of the distribution of fields at the muon site, and dynamic fluctuations, hence it is difficult to predict the expected behavior as done for LF \musr.
Despite this complexity, the change in behavior below $T^*$ suggests the energy scale set by the pseudogap is once again the important parameter, with the significant changes below $T^*$ likely occurring due to the fluctuating moment on the muon timescale.
Hence, our TF \musr\ results on \crtas\ are consistent with the pseudogap picture already discussed.

In conclusion, the results of electronic structure calculations provide an explanation of the low-$T$ transport and magnetic properties of half-metallic \crms\ (\meq); all published low-$T$ experimental observations can be explained from this electronic structure, often by considering electrons transitioning across the small pseudogap in the spin-up channel.
These effects do not require the inclusion of SOC in the calculation, and do not imply a change of DMI, or a change in the helical length.
We have measured the pseudogap in both materials using AC susceptibility, finding good agreement with DFT, and observe dynamic transitions across the gap occur over a wide range of fluctuation frequencies (0.1~Hz to MHz) observed with multiple techniques.
In \crnbs, \musr\ suggests that the CSL masks the low-$T$ effects.
The similar electronic and magnetic behavior of \crms\ (\meq) is consistent with the realization of a CSL in \crtas, a recent report~\cite{zhang2021chiral} has provided the first direct evidence of this.
Our work demonstrates that, even in materials where topological magnetism leads to desirable properties for applications, the magnetic properties cannot be considered in isolation, and a thorough understanding of the electronic structure is essential.

Part of this work was carried out at the STFC ISIS Facility, UK (\hyperlink{https://doi.org/10.5286/ISIS.E.RB2010327}{https://doi.org/10.5286/ISIS.E.RB2010327}), and S$\mu$S, Paul Scherrer Institut, Switzerland; we are grateful for the provision of beamtime.
We are grateful for computational support from both Durham Hamilton HPC and from the UK national high performance computing service, ARCHER, for which access was obtained via the UKCP consortium and funded by EPSRC (Grant No. EP/K013564/1).
The project was funded by EPSRC (UK) (Grant Nos.: EP/N032128/1 and EP/N024028/1).
M. N. Wilson acknowledges the support of the Natural Sciences and Engineering Research Council of Canada (NSERC).
Research data from this paper will be made available via Durham Collections at \textcolor{red}{XXX}.

\bibliography{bib}
	
\end{document}